# Phase-Only Holographic Assisted Planar Printing for Massively Multiplexed Optical Display and Encryption

*Yunxuan Wei*, *Tie Hu*, *Xing Feng*, *Ming Zhao\**, *Zhenyu Yang*

Y.X.Wei, T.Hu, X.Feng, M.Zhao, Z.Y.Yang

School of Optical and Electronic Information, Wuhan National Laboratory for Optoelectronics, Huazhong University of Science and Technology (HUST), Wuhan 430074, China

Y.X.Wei

Ming Hsieh Department of Electrical and Computer Engineering, University of Southern California, Los Angeles, California 90089, USA

E-mail: zhaoming@hust.edu.cn

Keywords: multiplexing, printings, holograms, optical encryption, metasurfaces

Multiplexed planar printings, made of single or few layer micro and nano optical platforms, are essential for high capacity display, information storage and encryption. Although having been developed rapidly, the demonstrated channels are still limited and also lack instantaneity. Here, holograms and printings, always regarded as two independent information coding domains with totally different principles, are combined together through our proposed angle multiplexing framework, leading to multiplexed printings with hundreds of channels. Based on such approach, we experimentally encode respectively 25 gray scale printings into 25 angles and even 8 gray scale videos into 8 angles with a phase-only spatial light modulator. As a bridge between printings and holograms, our method allows to generate printings combining various holographic methods. Beneficial from this, we demonstrate a gradient metasurface based 324 channel printing which multiplexes angles, polarizations and wavelengths simultaneously. Our work paves the way to flexibly angle-dependent printing display and massively multiplexed encryption systems.

## 1. Introduction

Planar optical platforms, including diffractive elements, liquid crystals and metasurfaces, are becoming ideal carriers for optical information because of the compact scale, the dens spatial sampling, the high efficiency, the easy fabrication and the flexible light parameter manipulation. Among them, printings and holograms are two dominant optical display and encryption medias, which are the reflectances of respectively the surface intensity and the near/far field diffraction.[1] As a result, most of past works on printings and holograms are relatively independent and their design principles also differ. Recently, effort has been made on integrating holograms and printings into one diffractive surface,[2-8] whereas the two medias are still treated as unrelated properties in different spatial domains.

Multiplexing, the fundation for switchable display, high capacity storage and encryption, is the core development for both printings and holograms. Compared to volume elements, planar optical devices only contain single or few layers, which limits the design freedom. In the last few decades, wavelength multiplexed,[9-12] polarization multiplexed,[13-18] space multiplexed,[19-22] spatial frequency multiplexed,[1] orbital angular momentum multiplexed[23-26] and hybrid multiplexed[27-30] printings and holograms using rich responsive metasurfaces and nano-structures[31-33] have experienced remarkable progresses. Recently, angular is also unlocked as a codable channel, by the coherent pixel design,[34] the complex amplitude encoding,[35, 36] and the structures with angle dependent optical responses.[37-42] However, among most of the above methods, especially for printings, the demonstrated channels still limit to several. Also, due to the lack of the instantaneity from micro and nano scale patterns, multiplexed printed video display is rarely achieved.

In this paper, to solve the above issues, we propose a massively multiplexed planar printing system named Holo-Printing which establishs the relation between printings and holograms. Such combination links the observed surface intensities with the frequency spectrums, so enabling highly flexible angle-dependent encoding (**Figure 1a**). With a phase-only spatial light modulator (SLM), we experimentally demonstrate a 25 channel Holo-Printing and discuss the possible applications in display and encryption. To show its potential in dynamic multiplexing, an 8 channel vedio display is shown through the same SLM. Moreover, because of the high compatibility of our method with holographic principles, many approaches in holography, like wavelength and polarization multiplexing, can be introduced to printings to expand the channel amount (Figure 1b). Based on this, we even realize a 324 channel Holo-Printing using a gradient metasurface.

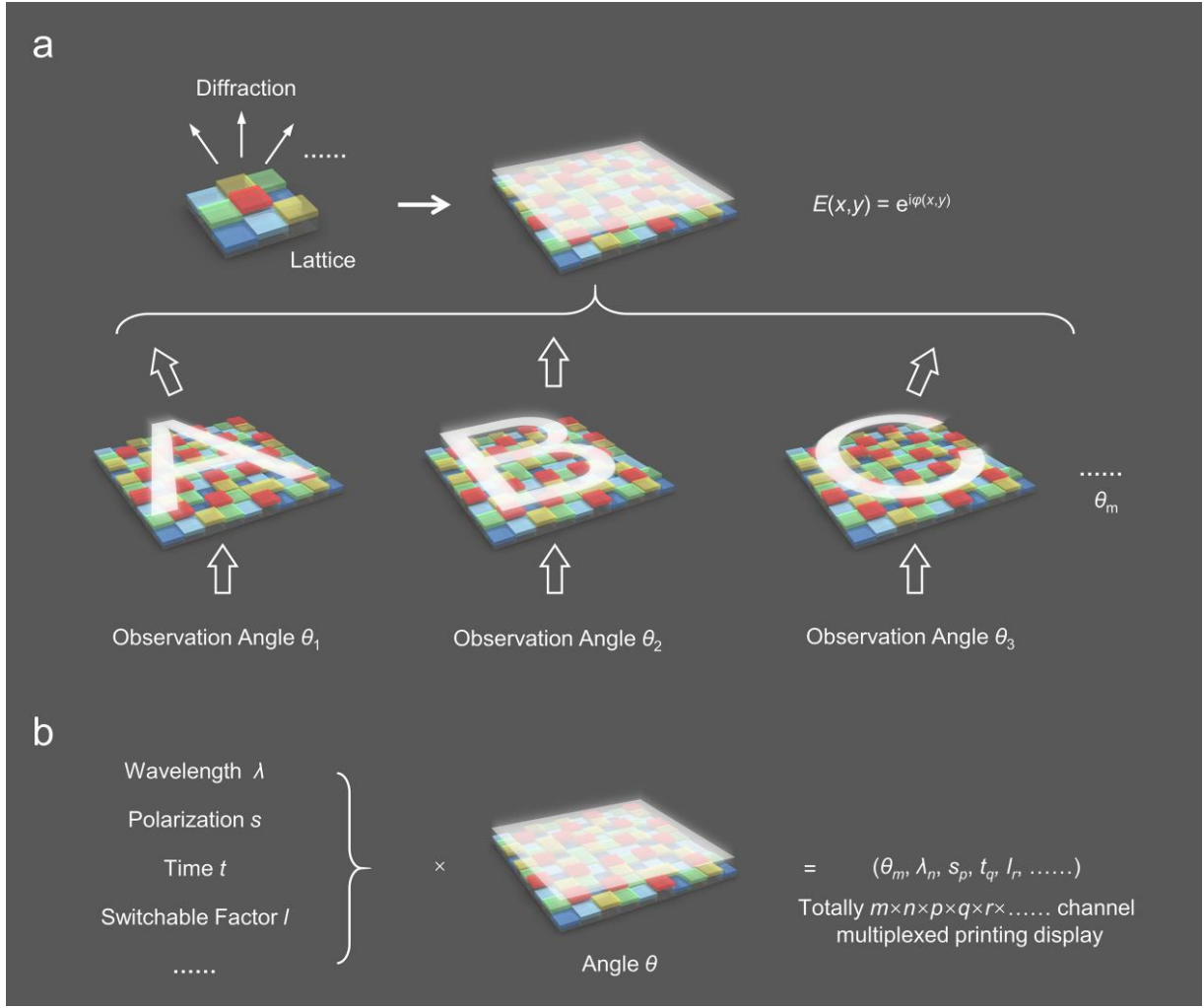

**Figure 1.** Concept of the Holo-Printing. a) Scheme of the working principle of the Holo-Printing made of phase-only diffractive lattices. When observed along different angles, it displays diverse printing images. b) By binding the Holo-Printing with switchable optical parameters through holograhic design approaches, a multidimensional printing system with massive channels is promising.

## 2. Results

### 2.1. Framework of Angle Multiplexing

Our angle multiplexing methd is analyzed based on rectangle shaped diffractive lattices (**Figure 2a**). For a periodic lattice with length $L_x$ and width $L_y$, when illuminated by normal incidence, its wave front can be decomposed into finite plane waves given by

$$E_0(x,y) = \sum_{m=-M}^{M} \sum_{n=-N}^{N} \exp(i2\pi(f_m x + f_n y)) \varepsilon_{mn}, \qquad f_m = \frac{m}{L_x}, \qquad f_n = \frac{n}{L_y} \#(1)$$

Where *m* and *n* are diffraction orders, *M* and *N* are their corresponding maximum values. $\mathcal{E}_{mn}$ is the complex amplitude of the (*m,n*) order plane wave. When such lattices, but with different $\mathcal{E}_{mn}$, are arranged together, the output field can be writen as

$$E(x,y) = \sum_{p=0}^{P-1}\sum_{q=0}^{Q-1} \text{rect}\left(\frac{x}{L_x} - p, \frac{y}{L_y} - q\right)\left(\sum_{m=-M}^{M}\sum_{n=-N}^{N} \exp(i2\pi(f_m x + f_n y))\mathcal{E}_{mn}^{pq}\right) \#(2)$$

Where (*p,q*) is the sequence of the lattice, *P* and *Q* are the maximum lattice amounts respectively in *x* and *y* directions. $\mathcal{E}_{mn}^{pq} = A_{mn}^{pq}\exp(i\varphi_{mn}^{pq})$ is the complex amplitude of the (*m,n*) order plane wave for lattice (*p,q*). Consider the total contribution of the (*m,n*) order component

$$E_{mn}(x,y) = \exp(i2\pi(f_m x + f_n y))\sum_{p=0}^{P-1}\sum_{q=0}^{Q-1}\text{rect}\left(\frac{x}{L_x} - p, \frac{y}{L_y} - q\right)\mathcal{E}_{mn}^{pq} \#(3)$$

Clearly, along the (*m,n*) order, the spatial light field is directly determined by $\mathcal{E}_{mn}^{pq}$. By carefully controlling this term, it is possible to realize angle dependent printings ($A_{mn}^{pq}$), phase elements ($\varphi_{mn}^{pq}$), and more complex devices. To explore the interference between different orders, $E_{mn}(x,y)$ is transformed into frequency domain

$$\tilde{E}_{mn}(f_x, f_y) = \delta(f_x - f_m, f_y - f_n)$$
$$* L_x L_y \underbrace{\text{sinc}(L_x f_x)\text{sinc}(L_y f_y)}_{\tilde{E}_1(f_x,f_y)} \underbrace{\sum_{p=0}^{P-1}\sum_{q=0}^{Q-1}\exp\left(-i2\pi(pL_x f_x + qL_y f_y)\right)\mathcal{E}_{mn}^{pq}}_{\tilde{E}_2(f_x,f_y)} \#(4)$$

Here, $\tilde{E}_2$ is a periodic term, which evenly spreads to the whole spectrum and causes the crosstalk between different orders. In contrast, $\tilde{E}_1$ is highly centralized and therefore can be used to reduce the crosstalk. Such reduction is proportional to the frequency gap between non-zero orders ($\mathcal{E}_{mn}^{pq} \neq 0$) due to the linewidth of $|\tilde{E}_1|^2$ (Figure 2b). To ensure that no overlap happens for the main lob of $|\tilde{E}_1|^2$, the minimum frequency distance between adjacent non-zero orders is

$$\Delta f_x = \frac{2}{L_x} = 2\Delta f_m, \quad \Delta f_y = \frac{2}{L_y} = 2\Delta f_n \#(5)$$

Therefore, by arranging the designed fields into different orders while ensuring a minimal interval of 1 order, every field will be produced in considerable fidelity. Under these conditions, the maximum amount of the multiplexed angle is up to (*M*+1)×(*N*+1), which reaches infinity when the diffractive surface is unbounded. We further detailed analyze the dependence of the crosstalk on the order interval and the frequency spectrum distribution in Section S1,2.

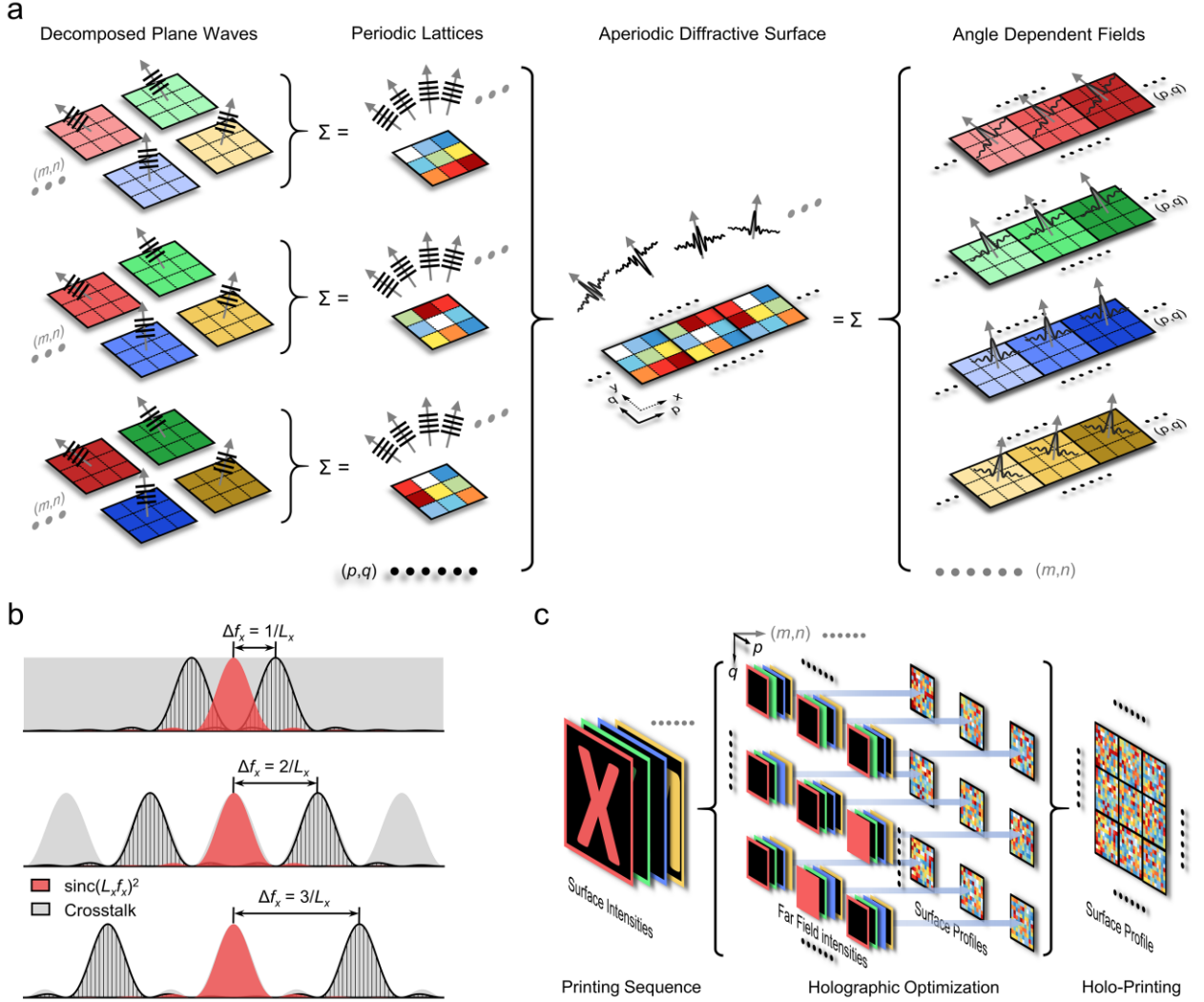

**Figure 2.** Framework of angle multiplexing and principle of the Holo-Printing. a) General design framework of angle multiplexed elements. b) Dependence of the crosstalk on the frequency gap between non-zero orders, along the $f_x$ axis. The red region and the black curve represent $|\tilde{E}_1|^2$ at respectively the target order and the adjacent non-zero orders. The gray patterns show the difference between target $|\tilde{E}_1|^2$ and the sum $|\tilde{E}_1|^2$ of infinite non-zero orders. c) General principle of the Holo-Printing. The surface printing sequence is converted into a Holo-Printing phase profile by holographic optimization.

## 2.2. Principle of the Holo-Printing

Following the framework above, the general design flow of the Holo-Printing is sketched in Figure 2c. The printing sequence determines the angle dependent amplitudes, in other words the amplitudes of the plane waves, of every lattice. Such plane waves are then refocused into corresponding lattices and give the wave front profiles. After all the lattices are traversed, they are finally assembled together to build up the whole Holo-Printing. Concequently, the key of this design is to achieve the desired plane wave amplitude $A_{mn}^{pq}$ for every lattice (*p*,*q*) and for every diffraction order (*m*,*n*) simultaneously. Here, inspired by computer generated holograms, we regard lattice (*p*,*q*) as an isolate hologram and generate the phase profile through the holographic optimization process. Although both complex amplitude and phase-only modulations are compatible for the Holo-Printing design, here we mainly consider the latter and employ the basic Gerchberg–Saxton (GS) algorithm to calculate the phase map.

For simplicity but without loss of generality, we discuss the detailed design and observation method of a 3×3 channel Holo-Printing comprising square lattices sized 11×11 pixels (**Figure 3a**). The field arrangement in frequency domain is also illustrated in Figure 3a, in which the red squares denote scattered unwanted light and the blue ones represent the printings. Besides the pattern we give here, the arrangement can be arbitrary determined as long as ensuring a least interval of one order. Figure 3b-f show the flow of encoding 9 pictures into the same surface made of 10×10 lattices. Before the loop, an intensity compensation process, depicted in Figure 3c, is added to maintain the gray scale of every image. The output intensity is set to uniformly unit 1 due to the phase-only modulation. During the compensation, we calculate the total intensity of the target images and choose the maximum as unit 1. Then the difference, between the total intensity and the output, is regarded as unwanted light and deflected to red squares. The phase pattern of a single lattice (Figure 3e) is generated by the GS algorithm. After walking through all the lattices, we get the phase of the Holo-Printing (Figure 3f).

As for the direct observation, the viewer should observe along the directions of the designed spatial frequencies, with a proper numerical aperture filtering out the undesired frequency components. Figure 3g draws the frequency spectrum of the Holo-Printing, clearly showing the separation of the printings. The solid rectangles circle the effective observation windows, corresponding to the regions of the main lobs. For smaller observation windows, the spectrum loss causes significant drop of the image quality (Section S3). Through inverse fast Fourier transform on the circled frequency components, the 9 printings are regenerated (Figure 3h), appearing as distinctive pictures of 9 numbers. The dot shaped profile of the lattice, resulting from the frequency cutoff, can be homogenized by enlarging the observation window at the expense of the larger crosstalk (Section S1). By centralizing the frequency spectrums of the printings, the crosstalk can be significantly reduced (Section S2). Outer the frequency region we care about, the high order diffractions also carry the printing information, the energy of which is however quite low (Section S4).

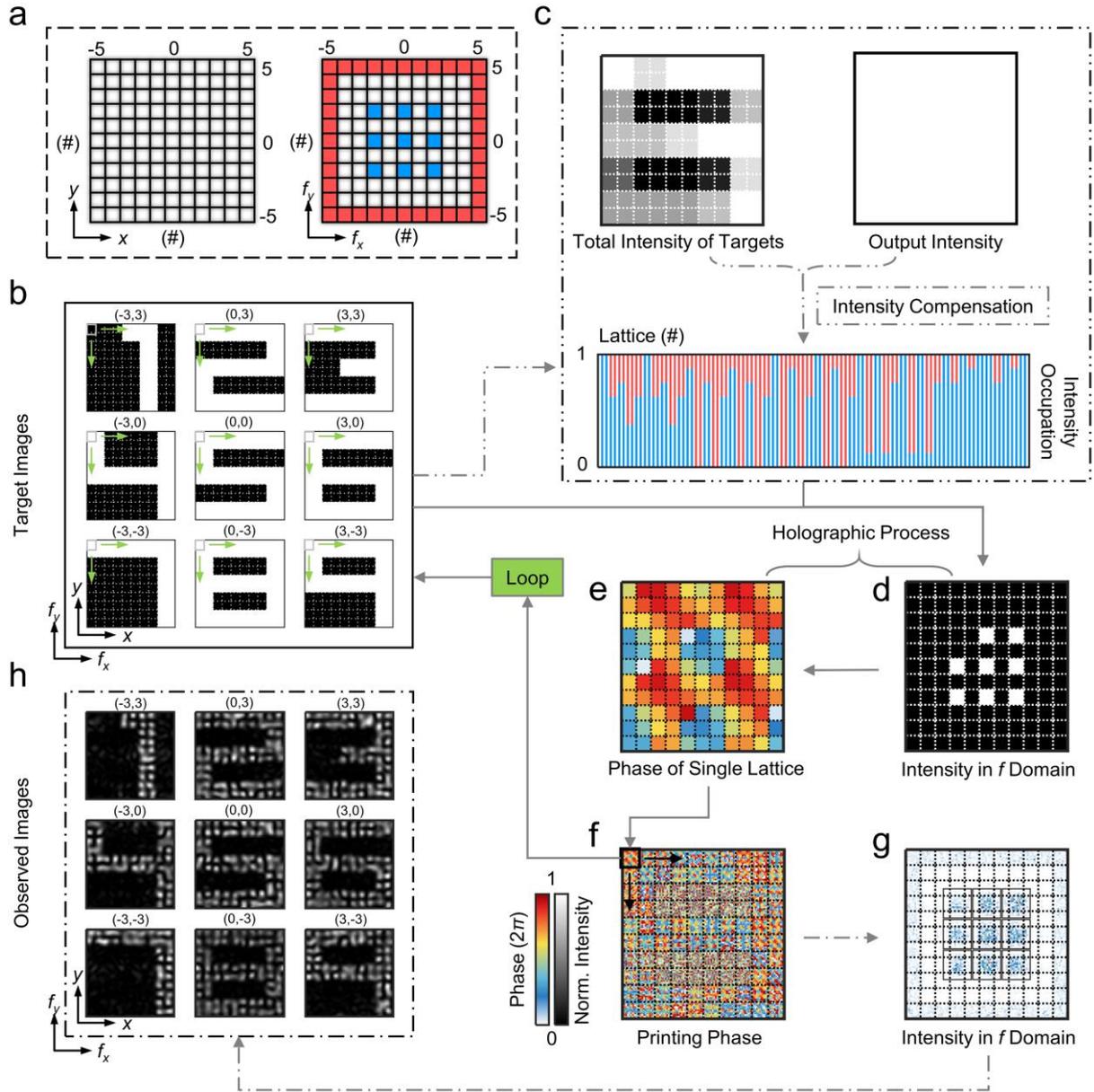

**Figure 3.** Design principle of the Holo-Printing. a) The square lattice (left) and frequency domain arrangement (right). b-f) Design flow of the Holo-Printing. The green squares in (b) and the black thick square in (f) denote the same lattice whose frequency spectrum and phase profile are in (d) and (e). The blue and red lines in (c) respectively represent the power ratio of the printings and the unwanted light. g) The frequency spectrum of the printing phase. Solid rectangles circle the main lob regions of the 9 printings. h) Observed printings calculated from corresponding frequency components marked in (g).

**2.3. Angle Multiplexed Holo-Printing Display and Encryption**

We then experimentally print a Holo-Printing encoded as many as 25 angle dependent printings using a phase-only SLM, and record the observations. Its frequency domain arrangement is drawn in **Figure 4a**, while the origin pictures, the phase profile, the calculated frequency spectrum and images are in Figure S11,12. For simplicity, we set up a 4*f* system to observe the SLM surface and use a circle aperture instead of a square hole as the frequency filter (see Section S5 for experiment setups). In Section S6, we show the impact of such filter replacement is minute on the image quality.

The direct imaging of the SLM surface and the frequency spectrum are shown in Figure 4b,c. As expected, the SLM equably reflects the illumination, making the printings invisible. At the spectrum plane, the desired diffraction pattern clearly appears. The brightest area is the superposition of the (0,0) printing and the zeroth order diffraction from the SLM. By moving the circular hole, we get the 25 coded printings displayed in Figure 4d. Except the one at (0,0) covered by the direct reflection, the remaining 24 printings are all legibly imaged. We attribute the overlap, of each two printings at orders symmetric to the origin (for example (+2, +4) and (-2, -4)), to the unideal diffraction efficiency of the SLM.

In addition to angle multiplexed display, the Holo-Printing is suitable for optical encryption, since the contained printing patterns are indistinguishable for the direct observation. Meantime, it allows variant intermediate coding keys, including observing angles, diffraction orders and imaging sequences, which can be translated into valid information only by decoding corresponding images from the Holo-Printing. For example, in Figure 4e, we show two keys respectively coded based on printing orders and angles, both obeying the rule of "Sign + Order/Angle". Although the two codes are totally different, after decoded using the Holo-Printing images in Figure 4d, they lead to the same sequence which provides far more complicated information.

Besides the frequency domain arrangement given above, we print the other Holo-Printing to demonstrate that the scattered light and the printings can be flexibly arranged. As shown in Figure 4f, 12 blue and 13 red pixels are randomly distributed in the central 9×9 area of the frequency spectrum, with an interval of 1 order between each other. The origin pictures, the phase map, the calculated frequency spectrum and printing images are collected in Figure S11,13. As for the experimental results, the directly reflected SLM pattern (Figure 4g) is still uniform, while the frequency spectrum components of the 25 encoded orders (Figure 4h) are clearly shown. As depicted in Figure 4i, after applying the frequency filter, 12 out of 25 orders display the printings of the 12 numbers from "one" to "twelve", while the other 13 ones appear as parasitic light. Such property can benefit display and encryption with angle selective visibility and increased security.

Apart from static printings, the Holo-Printing brings possibility for real-time multiplexed printing display. Here, we encode 8 directional sequences, each containing 60 printings, using 60 Holo-Printings. These Holo-Printings are played in about 6 seconds, corresponding to 8 channel videos with frame rates of around 10 frames per second (Movie S1). Figure 4j collects sample frames cut out from the recorded videos at the order (0,2) and (-2,0), which shows distinctive image sequences.

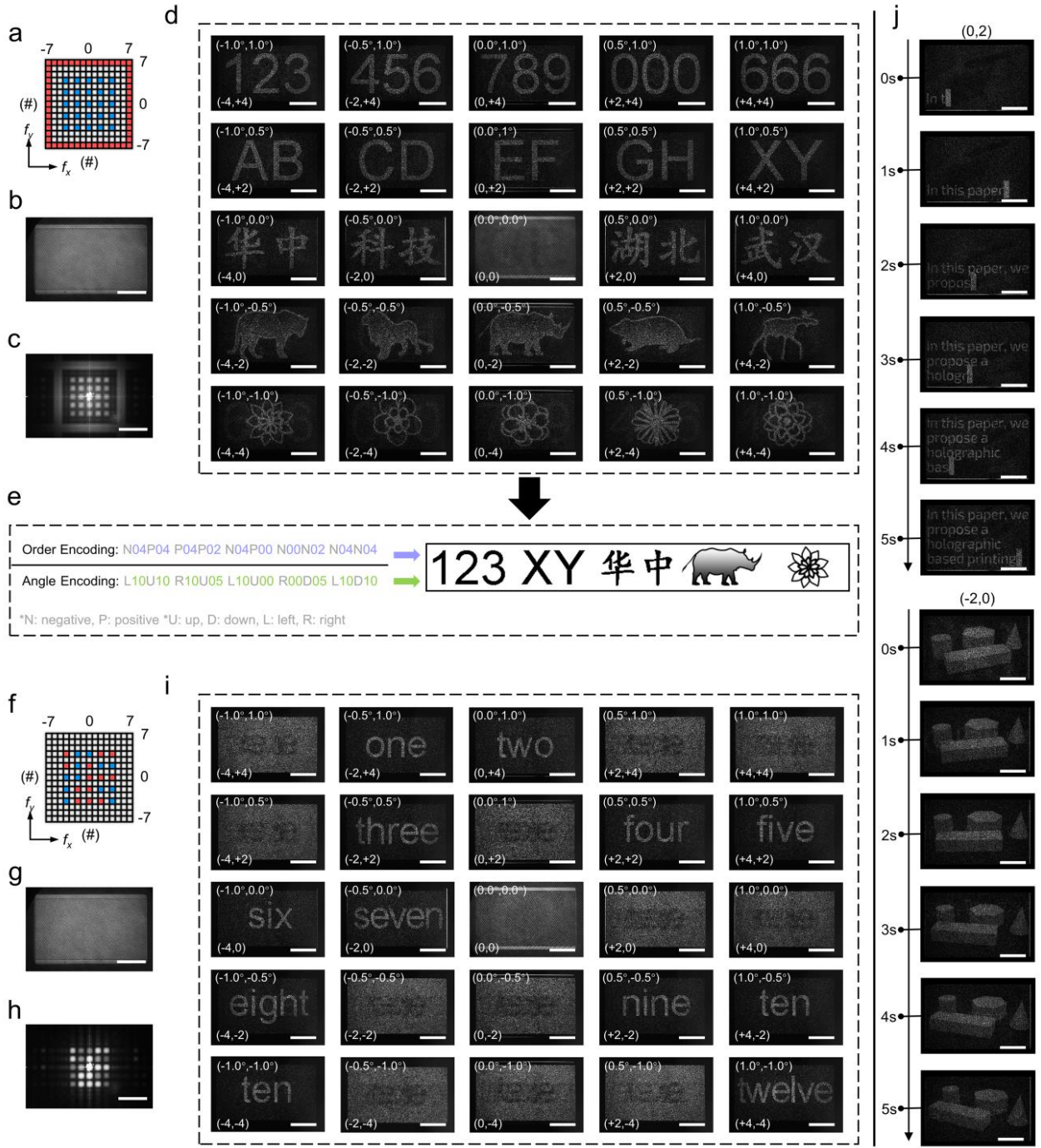

**Figure 4.** Experimental results of the angle multiplexed Holo-Printings. a-d) Design and observation of the first Holo-Printing, respectively showing (a) the frequency domain arrangement, (b) the surface of the SLM, (c) the frequency spectrum plane image, (d) the observed 5×5 multiplexed printings. Scale bars are 0.3 mm. e) An optical encryption example based on the first Holo-Printing. f-i) Design and observation of the second Holo-Printing, separately showing (f) the frequency domain arrangement, (g) the SLM surface, (h) the frequency spectrum plane image and (i) the 5×5 encoded images. The left bottom and the left top of each picture in (d,i) label the corresponding orders and the theoretical directions. j)

Sample frames cut out from the reorded angle multiplexed videos, respectively at the order (0,2) and (-2,0). Scale bars are 0.3 mm.

## 2.4. Massive Multiplexing with Extended Degrees of Design Freedom

Since only the phase is neccesasry for the Holo-Printing, many other design degrees, such as polarizations, wavelengths and spaces, are free for controlling and multiplexing. Among the planar optical platforms, the gradient metasurface is ideal for the multiple multiplexed Holo-Printing considering its subwavelength spatial resolution and flexible optical parameter manipulation. Here we utilize a bilayer gradient metasurface[11, 43] to simultaneously multiplex two wavelengths of 950 nm and 1550 nm and two polarization states of $x$ and $y$ linear polarizations. Such multiplexing leads to extra four states.

The bilayer metasurface unit, depicted in **Figure 5a**, is composed of four lower silicon nanopillars embedded in the square PDMS basement and an upper silicon nanopillar upon the basement. The large contrast of the lattice period results in that the phase shifts at the two wavelengths separately depend on the different layers (see Figure S14 for detailed optical responses). The selected units shown in Figure 5b, with different arm lengths $A_1$, $B_1$, $A_2$, $B_2$, allow arbitrary 8-level phase pairs for the four states (geometrical parameters and transmittances are collected in Figure S15). Then, four phase-only 9×9 channel Holo-Printings (frequency domain arrangements and origin pictures are in Figure S16) are encoded respectively into the four states, by a 648×1296 metasurface unit array (geometries in Figure S17,18). After constructed, the bilayer metasurface is simulated (the simulation method is in Section S7) and the 324 printing images are decoded numerically. Figure 5c draws the selected 100 ones, while the whole images are listed in Figure S19,20. As all of the printings are well displayed, the metasurface based Holo-Printing is proved to be able to significantly improve the multiplexing capacity of the present printing devices.

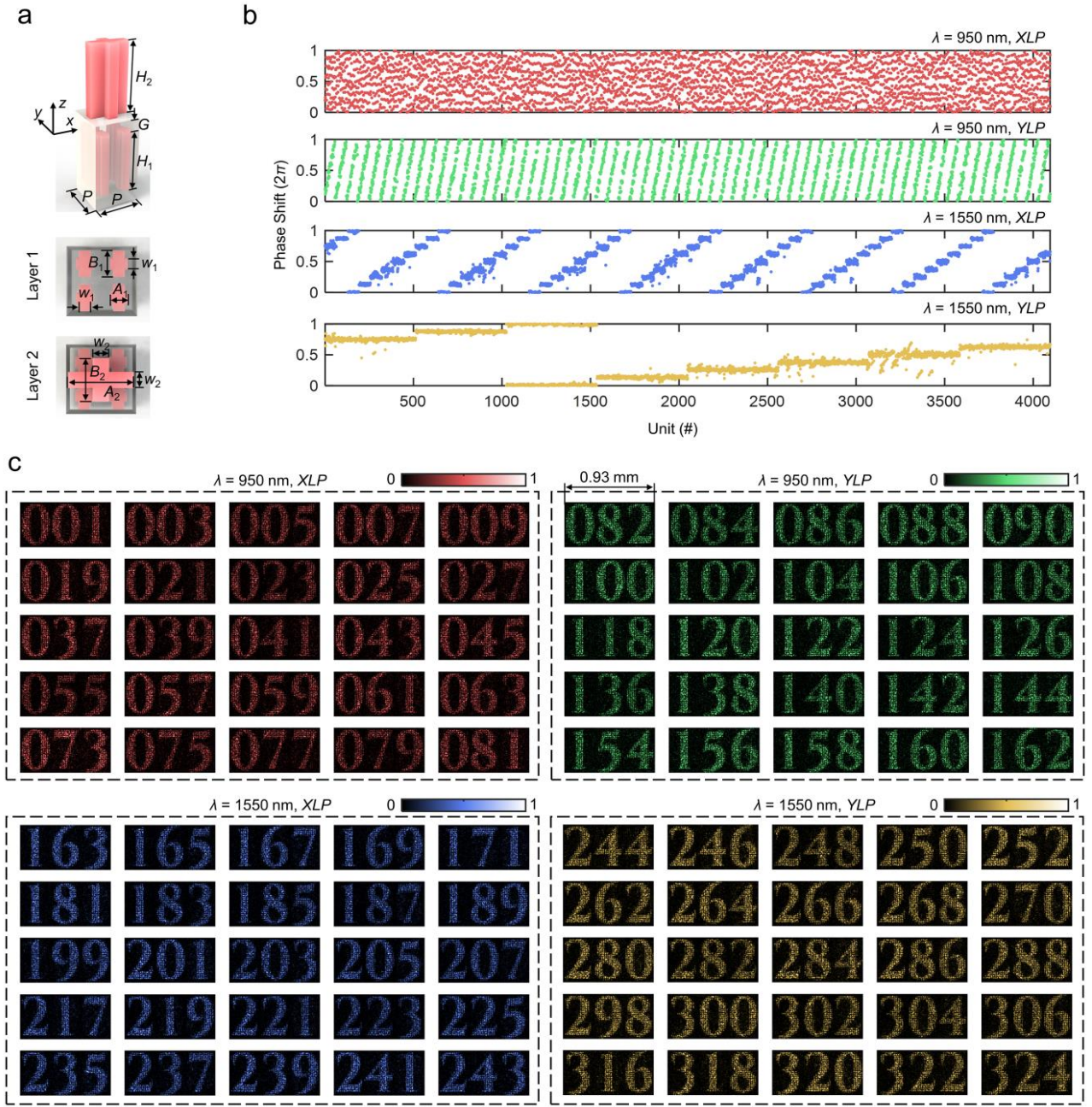

**Figure 5.** Results of the gradient metasurface based Holo-Printing. **a** Scheme of the metasurface unit, with fixed period $P$ = 720 nm, heights $H_1$ = 1000 nm and $H_2$ = 1100 nm, gap $G$ = 200 nm, and arm widths $w_1$ = 120 nm and $w_2$ = 150 nm. **b** Simulated phases of the selected 4096 units, at the two wavelengths and the two linear polarization states. **c** The selected 100 out of 324 calculated printing images.

## 3. Conclusion

To sum up, we have proposed a practical framework for optical angle multiplexing, based on which we illustrated the design method of the Holo-Printing. With a phase-only SLM, two tens channel angle multiplexed printings and an 8 channel video display were experimentally achieved. As an extension, we also demonstrated a 324 channel Holo-Printing using a wavelength and polarization multiplexed gradient metasurface. Beneficial from the intense channels and the real-time, the Holo-Printing is suitable for information storage and encryption and dynamic multiplexed displays. Particularly, our method introduces holographic design approaches into printings, therefore extends the flexibility and the potential advance of printing devices. By adding in more switchable factors, like phase change materials, the multiplexing capacity of a single Holo-Printing can be further increased.

**Supporting Information**

Supporting Information is available from the author.

**Acknowledgements**

Y.X.Wei, T.Hu and X.Feng contributed equally to this work.

**Funding.**

This work is supported by the Natural Science Foundation of China (No.62075073, 62135004 and 62075129), the Fundamental Research Funds for the Central Universities (No. 2019kfyXKJC038), State Key Laboratory of Advanced Optical Communication Systems and Networks, Shanghai Jiao Tong University (No. 2021GZKF007), and Key R & D project of Hubei Province (No. 2021BAA003).